\documentclass[aps,prl,superscriptaddress,showpacs,
preprintnumbers,amsmath,amssymb,amsfonts,floatfix,preprint]{revtex4-1}
\pdfoutput=1
\usepackage{graphicx}
\usepackage{color}
\usepackage{amsmath}
\usepackage[english]{babel}
\usepackage{amssymb}
\usepackage[normalem]{ulem}

\def\bc{\begin{center}} 
\def\ec{\end{center}}
\newcommand{\beq}{\begin{equation}}
\newcommand{\eneq}{\end{equation}}
\newcommand{\beqnn}{\begin{equation*}}
\newcommand{\eneqnn}{\end{equation*}}
\newcommand{\beqy}{\begin{eqnarray}}
\newcommand{\eneqy}{\end{eqnarray}}
\newcommand{\beqynn}{\begin{eqnarray*}}
\newcommand{\eneqynn}{\end{eqnarray*}}
\newcommand{\proj}[1]{\ket{#1}\bra{#1}}
\newcommand{\ket}[1]{\left |{#1}\right \rangle}

\newcommand{\bra}[1]{\langle #1 | }
\newcommand{\ave}[1]{\left[ #1 \right]_{\mathrm{ave}}}
\newcommand{\aveU}[1]{\left[ #1 \right]^U_{\mathrm{ave}}}
\newcommand{\aveHaar}[1]{\left[ #1 \right]^{\mathrm{Haar}}_{\mathrm{ave}}}

\newcommand{\qexps}[1]{\langle #1 \rangle}
\newcommand{\qexpl}[3]{\bra{#1} #2 \ket{#3}}

\newcommand{\Tr}{\mathrm{Tr}}

\begin{document}
\title{Pure state thermodynamics with matrix product states}
\author{Silvano Garnerone}		
\affiliation{Institute for Quantum Computing, University of Waterloo, Waterloo, ON N2L 3G1, Canada}

\begin{abstract}
We extend the formalism of pure state thermodynamics to matrix product states. In pure state thermodynamics finite temperature properties of quantum systems are derived without the need of statistical mechanics ensembles, but instead using typical properties of random pure states. We show that this formalism can be useful from the computational point of view when combined with tensor network algorithms. In particular, a recently introduced Monte Carlo algorithm is considered which samples matrix product states at random for the estimation of finite temperature observables. Here we characterize this algorithm as an $(\epsilon, \delta)$-approximation scheme and we analytically show that sampling one single state is sufficient to obtain a very good estimation of finite temperature expectation values. These results provide a substantial computational improvement with respect to similar algorithms for one-dimensional quantum systems based on uniformly distributed pure states. The analytical calculations are numerically supported simulating finite temperature interacting spin systems of size up to 100 qubits.  
\end{abstract}
%03.67.-a Quantum information
%05.30.-d Quantum statistical mechanic
%75.10.Jm Quantized spin models
%75.40.Mg Computer modeling and simulation of magnetic crical points
% 02.70.Uu applications of MC methods in mathematical physics
\maketitle

\section{Introduction}
Pure state thermodynamics explains the finite temperature behavior of sufficiently large quantum systems not in the standard ensemble framework of statistical mechanics, but instead considering typical properties of a single random quantum pure state. According to this picture statistical mechanics emerges as the result of extremely small statistical fluctuations in large enough closed quantum systems. Although it has been overlooked for a long time, this  approach can be traced back to the early days of quantum mechanics and to the study of thermodynamic properties in closed quantum systems \cite{Neumann2010,Goldstein2010}. Later on alternative derivations of pure state statistical mechanics can be found in Seth Lloyd's PhD thesis \cite{Lloyd2013}, and more recently in the literature on typicality \cite{Popescu2006,Goldstein2006a,reimann_foundation_2008}. The common feature behind  these works is the understanding that closed quantum systems described by pure states can behave, for many practical purposes, like statistical mechanic ensembles at equilibrium. From this perspective the effectiveness of ensembles in statistical mechanics is justified in view of the more fundamental quantum properties of the system. 

Until now most of the literature on pure state thermodynamics has focused on foundational aspects of quantum statistical mechanics \cite{gemmer2004quantum}, and on the explanation of the dynamics in experimentally realizable closed quantum systems or model Hamiltonians \cite{polkovnikov_colloquium:_2011}. Only recently it has been recognized that the mathematical formalism of quantum mechanics allows for the exploitation also at the computational level of pure state thermodynamics. In particular, the work of Sugiura and Shimizu \cite{Sugiura2012,Sugiura2013} quite remarkably shows that typicality can be used in numerical simulations to  approximate thermal quantum states; this is done sampling and properly manipulating one single pure state generated uniformly at random according to the Haar measure \cite{Haar1933}. Statistical properties of uniformly random pure states are such that the quality of the numerical approximation is extremely good and, for all practical purposes, one can simulate thermal quantum systems on $N$ qubits with a properly constructed random pure state of $N$ qubits (i.e. there is no need of additional degrees of freedom to purify the thermal state of the system). On the other hand it is well known that uniform random states are computationally hard to generate \cite{Poulin2011,Harrow2009,Hamma2012}, requiring an exponentially large number of parameters in system's size; hence the curse of dimensionality will restrict the use of algorithms like those in \cite{Sugiura2012,Sugiura2013} to systems of modest size in general. Here we address this problem, and we show that using an efficiently parametrizable class of states, known as Matrix Product States (MPS), it is still possible to simulate pure state thermodynamics at finite temperature with a polynomial amount of resources in system's size, provided there exists an MPS representation of the finite temperature state. The latter seems to be reasonable assumption in general, for not too small temperatures, according to results in \cite{Hastings2006}. More in particular, we describe a Monte Carlo (MC) procedure which samples from Random MPS (RMPS) \cite{Garnerone2010,Garnerone2010a,Garnerone2013}, and we characterize the algorithm's trade-off between accuracy and efficiency employing an $(\epsilon,\delta)$-approximation scheme. A side product of our investigation provides a result, of interest in the context of pseudo-random quantum circuits \cite{Emerson2003,Zanardi2013,Hamma2012,Hamma2012a}, showing that RMPS states are approximate 2-designs \cite{Harrow2009}.  
 
\section{Random MPS states} 
\begin{figure}%[!h]
\includegraphics[scale=0.35]{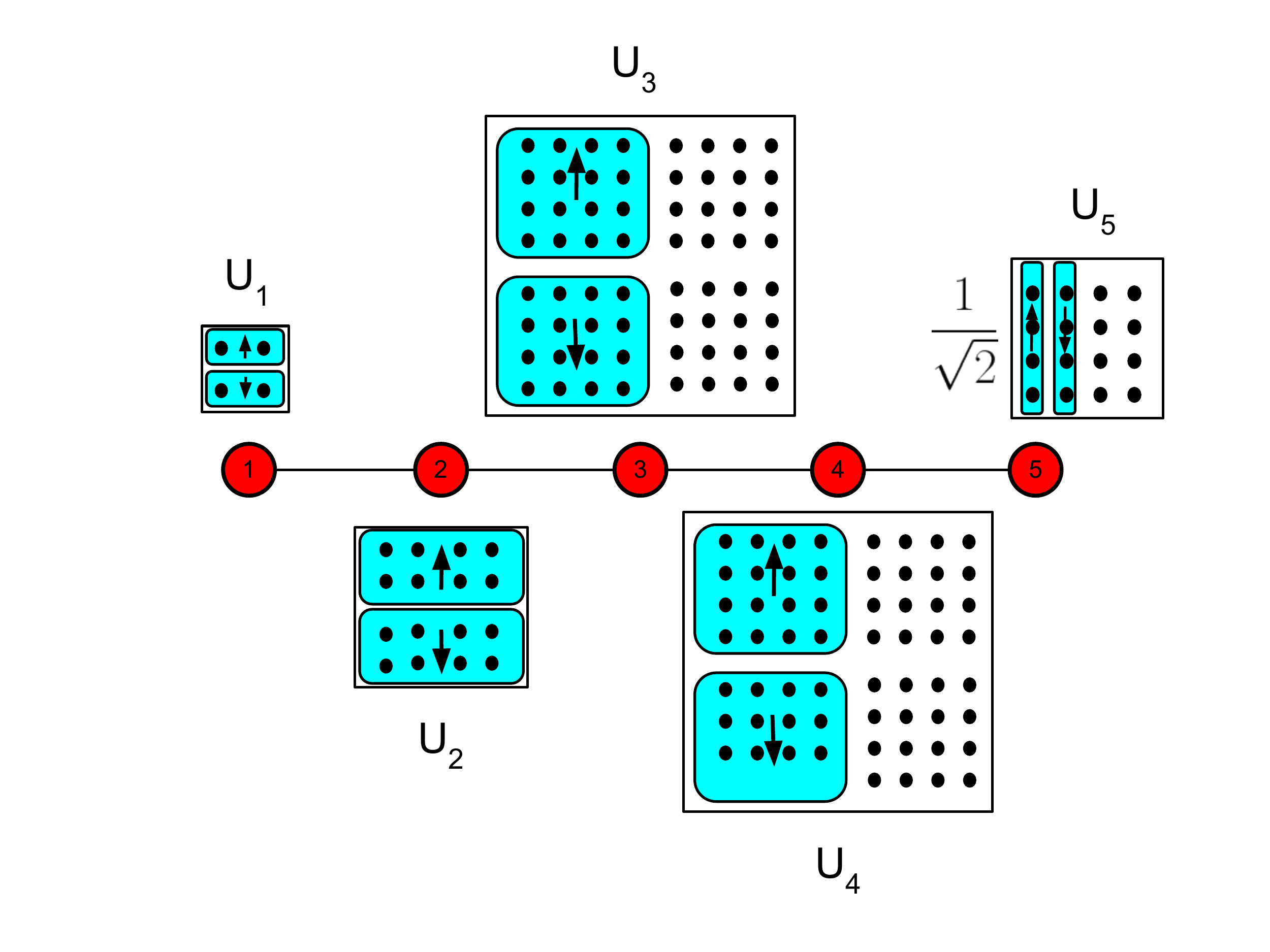} 
\caption{(Color online) An open chain of 5 qubits with the associated sequence of random unitaries generating the random MPS state; the $A$ matrices associated to the up or down spins are represented by sub-blocks inside each unitary matrix. }
\label{fig:figure1}
\end{figure}

We consider MPS with open boundary conditions associated to one-dimensional systems; generalizations to higher dimensions will be discussed in future works, while different boundary conditions can be dealt with the same formalism. In the following we use the standard notation for matrix product states, which is described for example in \cite{Schollwock2011}. An MPS $\ket{\psi}$ is completely characterized  by a set $\{ A^{\sigma_j}, j=1,\dots,N \}$ of matrices 
\beq
\ket{\psi}
=\sum_{\boldsymbol{\sigma},\bf{i}} 
A_{1,i_2}^{\sigma_1}A_{i_2,i_3}^{\sigma_2}\cdots 
A_{i_{N-2},i_{N-1}}^{\sigma_{N-1}}A_{i_{N-1},1}^{\sigma_{N}} \ket{\boldsymbol{\sigma}},
\label{eq:rmpsket} 
\eneq
with $\ket{\boldsymbol{\sigma}}\equiv \ket{\sigma_1 \sigma_2 \cdots \sigma_{N-1}\sigma_N}$ the computational basis. Open boundary conditions imply that $A^{\sigma_1}$ and $A^{\sigma_2}$ are respectively row and column vectors, while all other $A^{\sigma_j}$ are matrices whose greatest dimension is at most $\chi$, a parameter which is called the bond dimension of the MPS. We will often use the compact notation 
$
\ket{\psi}=A^{\sigma_1}A^{\sigma_2}\cdots 
A^{\sigma_{N-1}}A^{\sigma_{N}},
$
implicitly assuming the sum over  physical and auxiliary indices ($\boldsymbol{\sigma}$ and $\boldsymbol{i}$ respectively). It follows that for a chain of $N$ qubits an MPS is specified by no more than $2N \chi^2$ numbers which, for $\chi\sim \mathrm{poly}(N)$, is exponentially smaller than  $D \equiv 2^N$, the number of parameters required by a typical quantum state in the same Hilbert space $\mathcal{H}$. This compressed representation plays a key role in the efficiency of MPS algorithms \cite{Schollwock2011}.  
The ensemble of random MPS that we use in this work has been introduced in \cite{Garnerone2010,Garnerone2010a}, and used in \cite{Garnerone2013} to simulate quantum systems at finite temperatures in the microcanonical framework.  In the following we shortly summarize the construction of RMPS states \cite{a}. Consider a set of $N$ i.i.d. random unitary matrices, each one distributed according to the Haar measure, and with possibly different dimensions (see Fig.\ref{fig:figure1} for an example with five qubits). The $A^{\sigma_j=\{\uparrow,\downarrow\}}$ matrices defining the state are taken as sub-blocks of the random unitaries $U_{j}$: $A^{\sigma_j}\equiv L^{\sigma_j}U_{j}R_j$, where $L^{\sigma_j}$ and $R_j$ are truncation matrices selecting the proper sub-block in the unitary \cite{a}. The different dimensionality of the unitaries is a technical constraint implied by the normalization of the state, like the fact that the last unitary has to be divided by the square root of the local Hilbert space dimension \cite{a}. Although it might seem arbitrary, this construction is indeed related to a physical sequential generation of MPS states \cite{Schon2007}, and it inherits useful properties from the ensemble of random Haar unitaries. Given a proper sequence $U\equiv\{ U_j: j=1,\dots,N\}$ of i.i.d Haar unitaries we have all the ingredients necessary to specify the state in Eq.\ref{eq:rmpsket}.

In what follows we need the information provided by the first two moments of the ensemble of RMPS states: $\aveU{\psi}\equiv\aveU{\proj{\psi}}$ and $\aveU{\psi^{\otimes 2}}\equiv\aveU{\proj{\psi}^{\otimes 2}}$, where $\aveU{\cdot}$ denotes the average with respect to the set $U$ of unitaries. By using known properties of the Haar measure and the independence of the $U_j$'s at different sites $j$, it is possible to decompose the average over $U$ into a concatenated sequence of averages over single $U_j$'s, followed by contractions with neighbouring matrices \cite{a}
\beqynn
\aveU{\psi}
&=&
\left[A^{\sigma_1}\cdots
\left[A^{\sigma_{N-1}}
\left[ A^{\sigma_{N}} A^{\sigma'_N\dagger} \right]^{{U_{N}}}_{\mathrm{ave}}
%\right.\right.\\
%&&\times  
%\left.\left.
A^{\sigma'_{N-1}\dagger}\right]^{{U_{N-1}}}_{\mathrm{ave}}
\cdots A^{\sigma'_{1}\dagger}\right]^{{U_{1}}}_{\mathrm{ave}}\\
\aveU{\psi^{\otimes2}}&=&\left[B^{\sigma_1,\nu_1}\cdots
\left[ B^{\sigma_{N},\nu_{N}} B^{\sigma'_N,\nu'_N\dagger} \right]^{{U_{N}}}_{\mathrm{ave}}
\cdots B^{\sigma'_{1},\nu'_1\dagger}\right]^{{U_{1}}}_{\mathrm{ave}},
\eneqynn
where $B^{\sigma_j,\nu_j}\equiv A^{\sigma_j}\otimes A^{\nu_j}$. 
The calculation can be done exactly \cite{a} and the result is provided by 
\beqy
\aveU{{\psi}}&=&\aveHaar{{\psi}}=\frac{I}{D},\label{Eq:firstmom}\\
\aveU{{\psi}^{\otimes2}}&\approx &\aveHaar{{\psi}^{\otimes2}}
=
\frac{2\Pi^{\rm{sym}}}{D(D+1)}\label{Eq:secmom}
\eneqy
where we compare the RMPS result with the average with respect to uniformly distributed Haar states. $\Pi^{\rm{sym}}\equiv \frac{I+F}{2}$ is the projector over the symmetric subspace of $\mathcal{H}$ ($F$ is the swap operator and $I$ is the identity). The relative correction to $\aveHaar{{\psi}^{\otimes2}}$ in Eq.\ref{Eq:secmom} scales like
\beq\label{eq:relnorm}
\frac{\Vert \aveU{{\psi}^{\otimes2}} -  \aveHaar{{\psi}^{\otimes2}} \Vert_\infty}{\Vert\aveHaar{{\psi}^{\otimes2}}\Vert_\infty}
\sim O\left(\frac{1}{\chi}\right),
\eneq
where $\Vert \cdot \Vert_\infty$ is the operator norm (the largest singular value). This result can be understood intuitively, since for larger bond dimensions the RMPS ensemble spans a larger domain in $\mathcal{H}$. This implies that with larger $\chi$ we  should better approximate the Haar second-moment state, which is obtained considering the entire Hilbert space. By definition this also implies that RMPS states are approximate 2-designs \cite{Harrow2009}, where the approximation is controlled by the parameter $\chi$. In the next section we will use the information on the first two moments of the ensemble to characterize a Monte Carlo approximation scheme for the estimation of finite temperature expectation values. 

\section{$(\epsilon, \delta)$-approximation}
Suppose that we want to estimate a quantity Q, and that we have access to a stochastic device whose outcome is a random variable z with the properties that the mean Ave[z] is equal to Q, and the variance Var[z] is finite. We then use the stochastic device many times, assuming that the output of different trials are independent and identically distributed. Using Chebyshev's inequality one can show that after M trials we obtain an estimate of Q satisfying
\beq\label{eq:approxalgo}
\rm{Pr}\left[\Big{\lvert} \frac{\sum_{i=1}^M z_i}{M}-Q\Big{\rvert} \geq \epsilon Q\right]\leq \delta,
\eneq
with 
\beq\label{eq:delta}
\delta=\frac{{\rm Var[z]}}{{\rm Ave[z]^2}}\frac{1}{M\epsilon^2}.
\eneq
We have just described a $(\epsilon, \delta)$-approximation algorithm for evaluating Q, i.e. a Monte-Carlo algorithm that accepts as input an implicit description of Q together with two positive numbers $\epsilon$ and $\delta$, and it provides as output an estimate of Q satisfying Eq.\ref{eq:approxalgo}. 

For the problems we are interested in Q is given by the expectation value at finite temperature of an observable $B$
\beq
\qexps{B}_T\equiv\frac{\Tr{\rho_T B}}{\Tr{\rho_T}},
\eneq 
where $\rho_T$ could correspond for example to microcanonical or canonical mixed states
\beqy
\rho_{\rm{mic}}&\propto &\sum_{i\in\Delta E} \proj{E_i}\\
\rho_{\rm{can}}&\propto &e^{-\beta H},
\eneqy
and $\Delta E$ is a small energy window to which the eigenvalues of the energy eigenstates $\ket{E_i}$ belong; while $\beta$ and $H$ are respectively the inverse temperature and the Hamiltonian of the system. In \cite{Garnerone2013} the microcanonical setting was considered and a Monte Carlo  algorithm sampling random MPS has been provided which estimates the microcanonical expectation values of many-body systems. In what follows we derive rigorous bounds on the accuracy of the estimation obtained with such a MC algorithm. Note that these bounds are general and do not depend on the specific statistical ensemble. 

An effective way of approximately representing $\rho_{\rm mic}$ or $\rho_{\rm can}$ is through the iterative application of some operator $G$, which depends on the Hamiltonian $H$ of the system. As an example one can think of the familiar Trotter decomposition used in tDMRG \cite{Daley2004,White2004} or TEBD \cite{Vidal2003,Vidal2004} algorithms in imaginary time \cite{Schollwock2011}, where G is given by $exp(-\beta H/k)$ and $k$ is the number of times in which we have divided the interval $[0,\beta]$; while for the microcanonical ensemble one can use the procedure developed in \cite{Garnerone2013} [where G is given by $I-(H-E)^2/r^2$, for given scalar parameters $E$ and $r$]. For a given $G$ and using Eq.\ref{Eq:firstmom} it follows that, for  $k$ large enough, we can approximate the thermal mixed state as
\beq
\rho_T\propto  D \aveU{G^{\frac{k}{2}}\proj{\psi_i}G^{\frac{k}{2}}}=G^k.
\eneq
The above equation provides us with a strategy to sample RMPS states in order to estimate finite temperature expectation values. The quality of the approximation will be controlled by the second moment state through Chebyshev's inequality.

Defining $A\equiv G^\frac{k}{2}$, the random variable z$_i$ introduced at the beginning of this section is now given by  ${\rm z_i}\equiv{\qexpl{\psi_i}{ABA}{\psi_i}}/{\qexpl{\psi_i}{A^2}{\psi_i}}$. We want to estimate the accuracy of the following approximation 
\beq
\qexps{B}_T\approx\frac{1}{M}\sum_{i}^{M}{\rm z_i}.
\eneq
Since z$_i$ is the ratio of two random quantities, x$_i\equiv \qexpl{\psi_i}{ABA}{\psi_i}$ and y$_i\equiv  \qexpl{\psi_i}{A^2}{\psi_i}$, using standard error propagation and Eq.\ref{eq:relnorm} we can upper-bound the relative variance $\vert$Var[z]/Ave[z]$^2|$ with a function which goes to zero at least as $\chi^{-1}$ \cite{a}. Then from Eq.\ref{eq:approxalgo} and Eq.\ref{eq:delta} it follows that
\beq\label{eq:bound}
\rm{Pr}\left[\Big{\lvert} \frac{\sum_{i=1}^M z_i}{M}-{\rm Ave}[z]\Big{\rvert} \geq \epsilon {\rm Ave}[z]\right]\lesssim
\frac{1}{\chi}\frac{1}{M\epsilon^2},
\eneq
which implies that in order to have an $(\epsilon,\delta)$-approximation a number $M\sim (\delta \epsilon^2\chi)^{-1}$ of samplings is sufficient. Similarly one can also show that the relative error in approximating $\qexps{B}_T$ with  Ave[z] goes to zero at least as $\chi^{-1}$. These results show that pure state thermodynamics can indeed be simulated with a polynomial amount of resources in system's size. From a practical point of view we observe that the bound in Eq.\ref{eq:bound} is not tight and one can effectively use smaller bond-dimensions than those suggested by Eq.\ref{eq:bound} to obtain quantitatively accurate results. In the next section we will provide two simulations supporting the efficiency of the method. 

\begin{figure}%[!h]
\centerline{\includegraphics[scale=0.30]{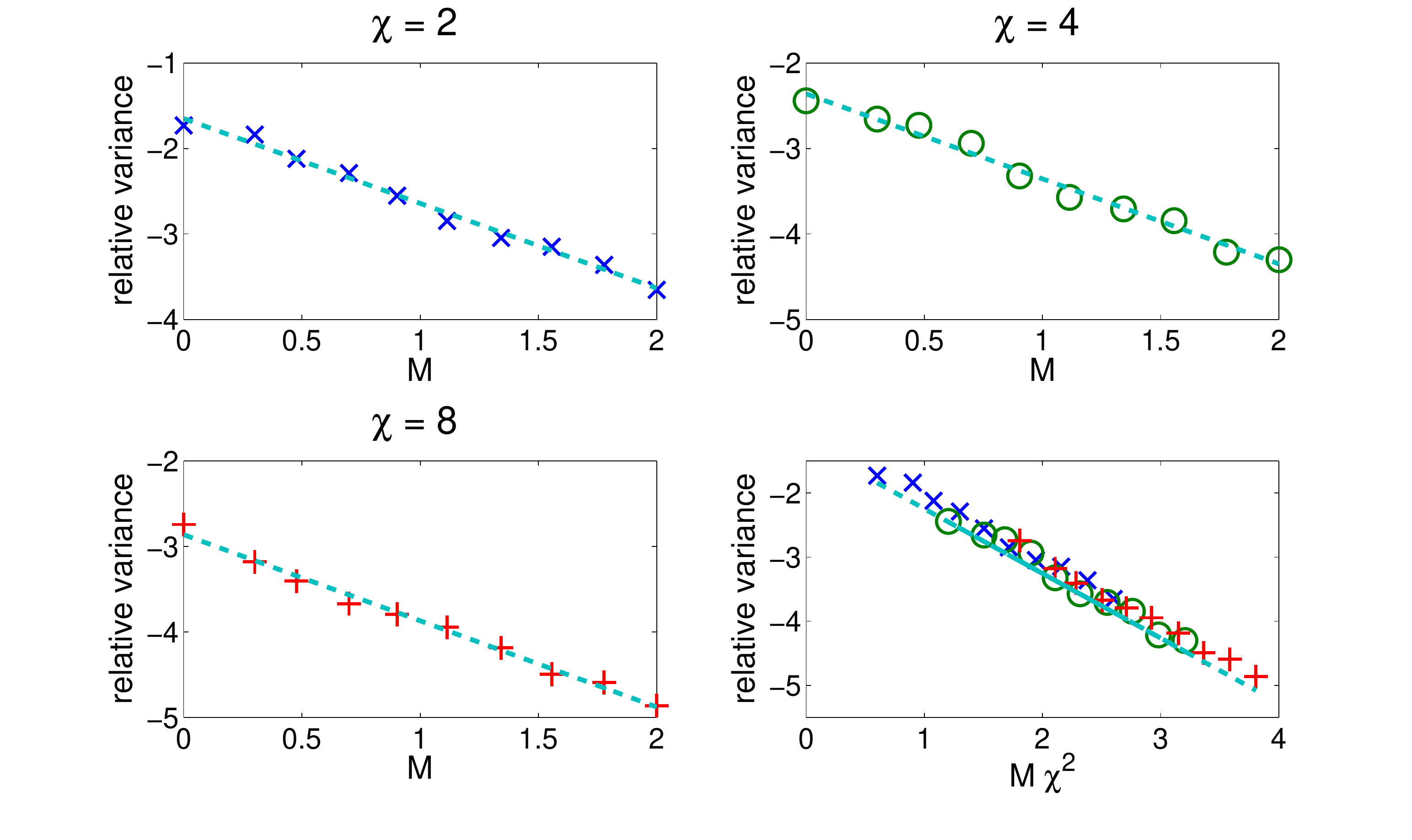}}
\caption{(Color online) The log-log plot of the relative variance of the random variable $\qexpl{\psi}{H^2}{\psi}$ (where $\ket{\psi}$ is a random MPS, and $H$ is a Ising Hamiltonian in a transverse field with $N=30$ and $\lambda=1.5$) as a function of the bond dimension $\chi$ and the number of samples $M$ for each run of the algorithm (over 100 runs in total). The bottom-right figure shows that curves with different bond dimension collapsed to a single one indicating a scaling function of the form $(M \chi^2)^{-1}$.}
\label{fig:figure2}
\end{figure}

\begin{figure}%[!h]
\centerline{\includegraphics[scale=0.30]{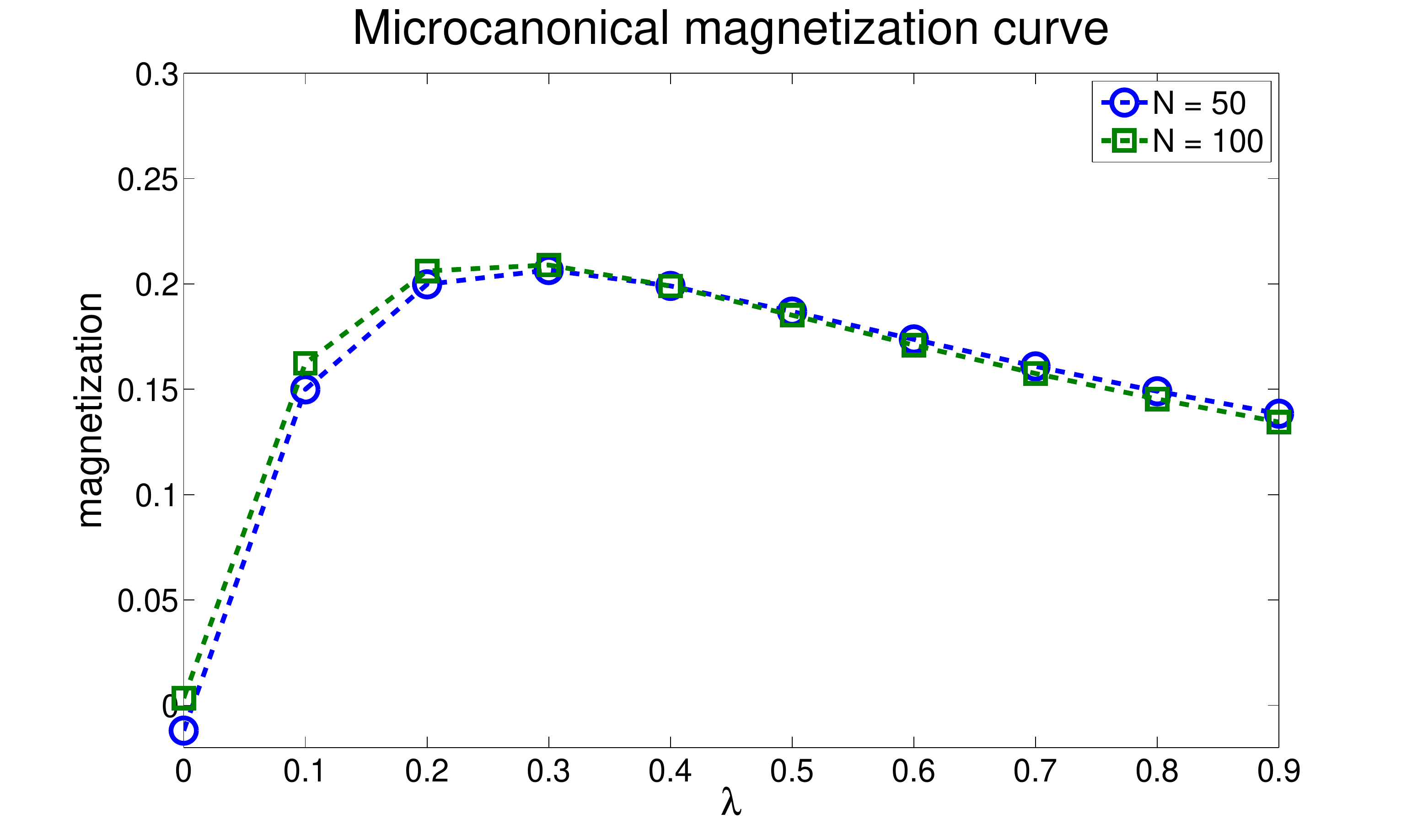}}
\caption{(Color online) The microcanonical magnetization curve of the Heisenberg chain in an external field $\lambda$. The two plots correspond to chains of 50 and 100 qubits, with an energy per spin equal to approximately $-0.15$. The data have been obtained sampling a single MPS of bond dimension $\chi=16$. The standard deviation obtained from 200 realizations of a RMPS state for each data point is comparable to the size of the circles and the squares in the plot.}
\label{fig:figure3}
\end{figure}

\section{Numerical experiments}
In the first numerical simulation we focus on the trade-off between the bond dimension $\chi$ and the number of samples in the MC estimate. This is a feature of the algorithm which is not only relevant for the estimation of finite temperature expectation values, but more in general for the estimation of the trace of an exponentially large operator. Consider the following Hamiltonian describing the Ising chain in transverse field
\beq
H=
\sum_{i=1}^{N-1}\sigma_i^x \sigma_{i+1}^x +\lambda \sum_{i=1}^N\sigma^z_i.
\eneq
For given parameters $N$ and $\lambda$ we want to estimate ${\rm x}\equiv\Tr{H^2}$ sampling MPS at random. Using Eq.\ref{Eq:firstmom} we need to evaluate the expectation value ${\rm x}_i\equiv\qexpl{\psi_i}{H^2}{\psi_i}$ with $M$ trials in order to estimate the exact result with $D \sum_i^M {\rm x}_i/M$. We are interested in particular on the scaling of the relative variance, provided by Var[x]/Ave[x]$^2$, as a function of the bond dimension $\chi$ and the number of trials $M$. In Fig.\ref{fig:figure2} we numerically estimate the relative variance for three different values of the bond-dimension $\chi={2,4,8}$. In the first three sub-figure we plot the relative variance increasing the number of sampling $M$ used in each run of the algorithm: i.e. for fixed $\chi$ and $M$, the algorithm is run 100 times to obtain the data point corresponding to the value of Var[x]/Ave[x]$^2$. The bottom-right sub-figure shows that the dependence of the three curves for different $\chi$ are consistently described by a function scaling like $(M\chi^2)^{-1}$. From the previous analytical estimate for the upper-bound on the fluctuations we would have expected a scaling like $(M\chi)^{-1}$. This suggests that for practical purposes relatively small bond-dimensions can be enough to provide very good estimates using Chebyshev's inequality. This numerical result supports the idea that RMPS states can be used in pure state thermodynamics as an efficient computational tool. Moreover, consistently with Eq.\ref{eq:bound}, one can trade larger bond dimensions for a smaller number of samples, or vice versa a larger number of samples for a smaller bond dimension. Since the sampling procedure can be easily parallelized the latter scheme would provide a faster way of obtaining the finite temperature result with the same accuracy. 

In the second numerical check we use the previous results to simulate at finite temperature a spin chain described by the Heisenberg Hamiltonian in an external field
\beq
H = - \sum_{i=1}^N \frac{1}{4}\left(\sigma^x_i \sigma^x_{i+1} + \sigma^y_i \sigma^y_{i+1} + \sigma^z_i \sigma^z_{i+1} \right) +\lambda \sigma^z_i. 
\eneq
We calculate the microcanonical magnetization curve (at the temperature corresponding to a microcanonical energy per spin $E/N=-0.15$) sampling one single  RMPS for chains of 50 and 100 qubits. The result of the simulation is shown in Fig.\ref{fig:figure3}, where one can see that the two curves (characterized by the same energy per spin) overlap very well, suggesting very small finite-size effects. Comparing this simulation with the one done in \cite{Garnerone2013} and \cite{Sugiura2012}, we can check that the statistical fluctuations (obtained with 200 samples) are smaller or equal to the size of the symbols used in Fig.\ref{fig:figure3}, supporting the result in Eq.\ref{eq:bound}.

\section{Conclusions}
We have provided an analytical estimate of the statical fluctuations in the evaluation of finite-temperature expectation values, induced by sampling random matrix product states. The results allow us to formulate a Monte Carlo $(\epsilon,\delta)$-approximation scheme which supports the use of random MPS states as a tool for pure state thermodynamics. We extend the results obtained in \cite{Sugiura2012,Sugiura2013} in the context of Haar distributed states, to a regime where the sampling procedure can be done efficiently, hence allowing for the study of much larger systems. Our findings are also of interest in the context of approximate state designs since the analytical evaluation of the second moment state shows that random MPS are approximate 2-designs, with a degree of approximation  controlled by an inverse polynomial function of the bond dimension. With respect to other algorithms for the simulation of finite-temperature systems \cite{Schollwock2011}, our scheme has the advantage of avoiding the introduction of auxiliary degrees of freedom, like ancillary qubits for the bath, which inevitably require additional computational resources. The Markov-Chain MC scheme proposed by White in \cite{white_minimally_2009} shares this same feature, although the Monte Carlo scheme discussed in the present work has the additional benefit of being easily parallelizable. For the future we plan to extend the study of random MPS to higher dimensional tensor networks, where the computational advantages provided by sampling random states could be even more significant. 

\bibliographystyle{apsrev4-1}
\bibliography{manuscript}

\appendix
\section{Supplemental Material}
\subsection{Normalized random MPS states}
The normalized random MPS is explicitly constructed as follows. Consider an open chain of $N$ qubits (the construction is easily generalizable to qudits, and chains with periodic boundary conditions) and fix a maximum bond-dimension $\chi$. Start from the left boundary qubit and generate a Haar-distributed 2 by 2 random unitary $U_1$. The left-boundary $A$ matrices are then defined by the two row vectors of $U_1$
\beqynn
A^{\sigma_1=\uparrow}[1:2]&\equiv & U_1[1,1:2],\\
A^{\sigma_1=\downarrow}[1:2]&\equiv & U_1[2,1:2]. 
\eneqynn
Consider now the second qubit and generate an independent 4-dimensional Haar-unitary matrix. Define the two $A$ matrices (of size 2 by 4) associated to the second qubits as follows 
\beqynn
A^{\sigma_2=\uparrow}[\cdot,\cdot]&\equiv & U_2[1:2,\cdot],\\
A^{\sigma_2=\downarrow}[\cdot,\cdot]&\equiv & U_2[3:4,\cdot]. 
\eneqynn 
Repeat this construction each time doubling the dimension of the random unitary, until one generates a $\chi$ by $\chi$ (with $\chi$ a power of 2) matrix $U_L$ at the $L$-th site
\beqynn
A^{\sigma_L=\uparrow}[\cdot,\cdot]&\equiv & U_L[1:\frac{\chi}{2},\cdot],\\
A^{\sigma_L=\downarrow}[\cdot,\cdot]&\equiv & U_L[\frac{\chi}{2}+1:\chi,\cdot]. 
\eneqynn 
For $j$ in between $L+1$ and $N-1$ generate i.i.d. $2\chi$ by $2\chi$ Haar-unitaries $U_j$, and define the corresponding $A^{\sigma_j}$ matrices as follows
\beqynn
A^{\sigma_j=\uparrow}[\cdot,\cdot]&\equiv & U_j[1:\chi,\cdot],\\
A^{\sigma_j=\downarrow}[\cdot,\cdot]&\equiv & U_j[\chi+1:2\chi,\cdot]. 
\eneqynn 
Finally consider the last site, for which a $\chi$ by $\chi$ Haar-unitary has to be generated and the $A^{\sigma_N}$ column vectors will be given by  
\beqynn
A^{\sigma_N=\uparrow}[1:\chi]&\equiv & \frac{U_N[1:\chi,1]}{\sqrt{2}},\\
A^{\sigma_N=\downarrow}[1:\chi]&\equiv & \frac{U_N[1:\chi,2]}{\sqrt{2}}. 
\eneqynn 
Note that we rescale the last A-matrix with the square root of the local Hilbert space dimension in order to have a normalized random MPS state (this step will be clear later when evaluating the norm of the state). 

At each step the $A^{\sigma_j}$ matrix can be written in the form $$A^{\sigma_j}=L^{\sigma_j}U_jR_j,$$ with $L^{\sigma_j}$ and $R_j$ proper truncation matrices 
\beqynn
L^{\sigma_j=\uparrow}&\equiv&\left[I; \boldsymbol{0} \right],\\
L^{\sigma_j=\downarrow}&\equiv&\left[\boldsymbol{0};I \right],\\
R_j&=&\left\{ I\;\rm{or}\; 
\left[
\begin{array}{c}
I \\ 
\boldsymbol{0}
\end{array} 
\right] 
\right\},
\eneqynn
where $\boldsymbol{0}$ is an array of zeros and $I$ the identity matrix. The dimension of the specific truncation matrices are fully specified by the sequential construction (see also Fig.1 in the main manuscript for a cartoon representation in the case of 5 qubits). 

The normalized state
\beq
\ket{\psi}=A^{\sigma_1}A^{\sigma_2}\cdots 
A^{\sigma_{N-1}}A^{\sigma_{N}}
\eneq
is then consistently defined (we implicitly assume the sum over physical and auxiliary indices). The above construction implies canonical left-normalized states such that for any $l\in\{1,\cdots,N\}$
\beq
\sum_{\sigma_l=\{\uparrow,\downarrow\}}A^{\dagger \sigma_l}A^{\sigma_l}=I.
\eneq
The overlap $\langle \psi \ket{\psi}$ for a fix random state is given by the following sequence of iterative contractions 
\beqynn
\langle \psi \ket{\psi}&=&
A^{\sigma'_N\dagger}A^{\sigma'_{N-1}\dagger}\cdots 
A^{\sigma'_{2}\dagger}
\left(A^{\sigma'_{1}\dagger}
A^{\sigma_1}\right)
A^{\sigma_2}\cdots 
A^{\sigma_{N-1}}A^{\sigma_{N}}\\
&=&
A^{\sigma'_N\dagger}A^{\sigma'_{N-1}\dagger}\cdots 
A^{\sigma'_{2}\dagger}
\left(
R^\dagger_1U^\dagger_1L^{\sigma'_1\dagger}
L^{\sigma_1}U_1R_1
\right)
A^{\sigma_2}\cdots 
A^{\sigma_{N-1}}A^{\sigma_{N}}\\
&=&
A^{\sigma'_N\dagger}A^{\sigma'_{N-1}\dagger}\cdots 
A^{\sigma'_{2}\dagger}
\left(R^\dagger_1U^\dagger_1
I
U_1R_1
\right)
A^{\sigma_2}\cdots 
A^{\sigma_{N-1}}A^{\sigma_{N}}\\
&=&
A^{\sigma'_N\dagger}A^{\sigma'_{N-1}\dagger}\cdots 
A^{\sigma'_{2}\dagger}
\left(R^\dagger_1
I
R_1
\right)
A^{\sigma_2}\cdots 
A^{\sigma_{N-1}}A^{\sigma_{N}}\\
&=&
A^{\sigma'_N\dagger}A^{\sigma'_{N-1}\dagger}\cdots 
\left(A^{\sigma'_{2}\dagger}
I
A^{\sigma_2}
\right)
\cdots 
A^{\sigma_{N-1}}A^{\sigma_{N}}\\
&=&
A^{\sigma'_N\dagger}A^{\sigma'_{N-1}\dagger}\cdots 
\left(R^\dagger_2U^\dagger_2L^{\sigma'_2\dagger}
L^{\sigma_2}U_2R_2
\right)
\cdots 
A^{\sigma_{N-1}}A^{\sigma_{N}}\\
&=&A^{\sigma'_N\dagger}A^{\sigma'_{N-1}\dagger}\cdots 
\left(I\right)
\cdots 
A^{\sigma_{N-1}}A^{\sigma_{N}}\\
&=&\cdots\\
&=&A^{\sigma'_N\dagger}
\left(A^{\sigma'_{N-1}\dagger}
I
A^{\sigma_{N-1}}\right)
A^{\sigma_{N}}\\
&=&A^{\sigma'_N\dagger}
\left(R^\dagger_{N-1}U^\dagger_{N-1}L^{\sigma'_{N-1}\dagger}
L^{\sigma_{N-1}}U_{N-1}R_{N-1}\right)
A^{\sigma_{N}}\\
&=&A^{\sigma'_N\dagger}
I
A^{\sigma_{N}}\\
&=&
R^\dagger_{N}U^\dagger_{N}L^{\sigma'_{N}\dagger}
L^{\sigma_{N}}U_{N}R_{N}\\
&=&A^{*\sigma_N}_{1,\cdot}A^{\sigma_N}_{\cdot,1}=1,
\eneqynn 
where each step follows from the previous definitions.

\subsection{First moment state}
In this section we detail the steps needed to evaluate the average state of the ensemble (which is already discussed in Ref.\cite{Garnerone2010a,Garnerone2010}): 
\beqy 
\aveU{\proj{\psi}}
&=& 
\aveU{A^{\sigma_1}\cdots A^{\sigma_{N}} A^{\sigma'_N\dagger}\cdots A^{\sigma'_{1}\dagger}}\\
&=&\left[A^{\sigma_1}\cdots
\left[A^{\sigma_{N-1}}
\left[ A^{\sigma_{N}} A^{\sigma'_N\dagger} \right]^{\mathrm{U_{N}}}_{\mathrm{ave}}
A^{\sigma'_{N-1}\dagger}\right]^{\mathrm{U_{N-1}}}_{\mathrm{ave}}
\cdots A^{\sigma'_{1}\dagger}\right]^{\mathrm{U_{1}}}_{\mathrm{ave}},
\label{eq:aveiter} 
\eneqy
where we just used the independence of the unitaries at each site of the chain to rewrite the total average over the sequence $U$ as a concatenation of averages over single $U_j$ unitaries. Then the problem is reduced to the evaluation of averages at each site, which can be done using the following expressions
\beqy
\left[ A^{\sigma_{j}} A^{\sigma'_j\dagger} \right]^{\mathrm{U_{j}}}_{\mathrm{ave}}
&=&\left[ L^{\sigma_j}U_j R_j 
R_j^\dagger U_j^\dagger L^{\sigma'_j\dagger} \right]^{\mathrm{U_j}}_{\mathrm{ave}}\nonumber\\
&=&L^{\sigma_j}\left[U_j R_j R_j^\dagger U_j^\dagger \right]^{\mathrm{U_j}}_{\mathrm{ave}}L^{\sigma'_j\dagger}\\ 
&=&\frac{\Tr{(R_j R_j^\dagger)}}{{\rm dim}(U_j)}L^{\sigma_j}{I}L^{\sigma'_j\dagger}
=\frac{\Tr{(R_j R_j^\dagger)}}{{\rm dim}(U_j)}\left(L^{\uparrow}L^{\uparrow\dagger}+L^{\downarrow}L^{\downarrow\dagger}\right)\\
&=&\frac{\Tr{(R_j R_j^\dagger)}}{{\rm dim}(U_j)}I,
\eneqy
where we made use of the twirling identity and the definition of the $L$ matrices. The concatenation of the averages leads  to the following expression
\beq\label{eq:avestate}
\frac{\chi_2}{d\chi_1}\frac{\chi_3}{d\chi_2}\cdots\frac{\chi_{N-1}}{d\chi_{N-2}}\frac{I}{d\chi_{N-1}}
=\frac{I}{d^{N-1}\chi_1}=\frac{I}{d^N}=\frac{I}{D},
\eneq 
where $\chi_j$ is the bond dimension at site $j$ in the chain. Note that the above result is independent of the maximum bond dimension $\chi$ of the state. 
\subsection{Second moment state}
In this section we provide the details of the derivation of the second moment state. Since this calculation is quite lengthy in the general case, we will first derive the explicit result for the simple case of chain with $2$ qubits in order to illustrate the structure of the calculation and of the solution, then we will obtain the general expression. 

For two qubits the second moment state is given by
\beqy 
\aveU{\proj{\psi}\otimes \proj{\psi}}=\sum_{\boldsymbol{\sigma,\sigma',\nu,\nu'}}
\left[
A^{\sigma_1} A^{\sigma_{2}}
A^{\nu_1} A^{\nu_{2}} 
A^{\sigma'_2\dagger} A^{\sigma'_{1}\dagger}
A^{\nu'_2\dagger} A^{\nu'_{1}\dagger}\right]^\mathrm{U}_\mathrm{ave}
\ket{\boldsymbol{\sigma\,\nu}}\bra{\boldsymbol{\sigma'\,\nu'}}.
\eneqy  
We use a well known property of the Haar measure, namely that the average rotation of an operator $O$ with respect to a random unitary tensored twice is provided by
\beqynn
\aveU{
(U\otimes U)O(U \otimes U)^\dagger 
}
=\Pi_{s}\frac{\Tr{(O \Pi_{s})}}{\mathrm{dim}(\Pi_s)}
+\Pi_{a}\frac{\Tr{(O \Pi_{a})}}{\mathrm{dim}(\Pi_a)},
\eneqynn
with $$\Pi_{s}=\frac{{I}+ F}{2},$$ and $$\Pi_{a}=\frac{{I}- F}{2},$$ ${F}$ being the swap operator acting on the computational basis 
as $${F}\ket{\sigma_i \nu_i}\equiv\ket{\nu_i \sigma_i}.$$
The average of the $2$ qubit chain can be performed in two steps as follows (here we write the matrices' components in order to make clear the structure of the contractions, and the asterisk denotes matrix conjugation)
\beqynn
&&\aveU{
A_{1,i_1}^{\sigma_1}A_{1,k_1}^{\nu_1}
A_{i_1,1}^{\sigma_2}A_{k_1,1}^{\nu_2} 
A_{1,j_1}^{\sigma'_2*}A_{1,l_1}^{\nu'_2*}
A_{j_1,1}^{\sigma'_1*}A_{l_1,1}^{\nu'_1*}
}\\
&=&\ave{
A_{1,i_1}^{\sigma_1}A_{1,k_1}^{\nu_1}
\ave{A_{i_1,1}^{\sigma_2}A_{k_1,1}^{\nu_2} 
A_{1,j_1}^{\sigma'_2*}A_{1,l_1}^{\nu'_2*}}^{U_2}
A_{j_1,1}^{\sigma'_1*}A_{l_1,1}^{\nu'_1*}
}^{U_1}.
\eneqynn
Consider the internal average over $U_2$, using the previous result on the Haar measure we have
\beqynn
\ave{A_{i_1,1}^{\sigma_2}A_{k_1,1}^{\nu_2} 
A_{1,j_1}^{\sigma'_2*}A_{1,l_1}^{\nu'_2*}}^{U_2}
&=&\ave{U_2^{\otimes 2}\ket{\sigma_2 \nu_2}\bra{\sigma'_2 \nu'_2}
U_2^{\dagger \otimes 2}}^{U_2}\\
&=&
\Pi_{s}\frac{\qexpl{\sigma'_2 \nu'_2}{\Pi_{s}}{\sigma_2 \nu_2}}{d^2\mathrm{dim}(\Pi_s)}
+\Pi_{a}\frac{\qexpl{\sigma'_2 \nu'_2}{\Pi_{a}}{\sigma_2 \nu_2}}{d^2\mathrm{dim}(\Pi_a)}.
\eneqynn
Each symmetric or anti-symmetric projector, contracted with the matrices of the first qubit, provides
\beqynn
\ave{
A_{1,i_1}^{\sigma_1}A_{1,k_1}^{\nu_1}
\Pi_{s,a}
A_{j_1,1}^{*\sigma'_1}A_{l_1,1}^{*\nu'_1}}^{U_1}
=\bra{\sigma_1\nu_1}
\ave{U_1^{\otimes 2}\Pi_{s,a}U_1^{\dagger\otimes 2}}^{U_1}
\ket{\sigma'_1\nu'_1}
=\bra{\sigma_1\nu_1}\Pi_{s,a}\ket{\sigma'_1\nu'_1},
\eneqynn
where we used the fact the projectors commutes with $U_1^{\otimes 2}$. Putting everything together we have 
\beqynn
\aveU{\psi^{\otimes 2}}=\frac{\Pi_s\otimes\Pi_s}{d^2\mathrm{dim}(\Pi_s)}+
\frac{\Pi_a\otimes\Pi_a}{d^2\mathrm{dim}(\Pi_a)}.
\eneqynn

In the case of more than 2 qubits we have to repeat the sequence of averages and contractions in a concatenated way, similarly to what has been done for the evaluation of the first moment state. To simplify notation let us define $D_s^\chi \equiv \chi(\chi+1)/2$ and $D_a^\chi \equiv \chi(\chi-1)/2$. Then let's consider the outcome (we denote it as $\rho_R$) of the first two most internal averages [those with respect to the $N$-th and $(N-1)$-th qubits], which following the above two-qubits calculation provides 
\beq
\rho_R=\frac{(I+F)\otimes \Pi^{phys}_s}{2d^2 D_s^{\chi}}
+\frac{(I-F)\otimes \Pi^{phys}_a}{2d^2 D_a^{\chi}},
\eneq
where the superscript $phys$ denotes an operator acting only on the physical space (in this case the one associated to the last qubit), while $I$ and $F$ act on the auxiliary space associated to the bond dimension. To proceed with the calculation we now need to update $\rho_R$ contracting it with the $A$ matrices associated to the $(N-2)$-th qubit
\beq
\rho_R
\leftarrow 
(L^{\sigma_{N-2}}\otimes L^{\nu_{N-2}})(UR)^{\otimes 2}
\rho_R
(UR)^{\dagger \otimes 2}
(L^{\nu'_{N-2}\dagger}\otimes L^{\sigma'_{N-2}\dagger}).
\eneq
The contraction will affect only the operators acting on the auxiliary space. Defining $$\alpha\equiv\frac{D_s^{\chi}}{D_s^{2\chi}}=\frac{\chi+1}{2(2\chi+1)}$$ and $$\beta\equiv\frac{D_a^{\chi}}{D_a^{2\chi}}=\frac{\chi-1}{2(2\chi-1)},$$ one can easily check the following identities
\beq
(UR)^{\otimes 2}I(UR)^{\dagger \otimes 2}=
\alpha \Pi_s + \beta \Pi_a
\eneq
\beq
(L^{\sigma_j}\otimes L^{\nu_j})(\alpha \Pi_s + \beta \Pi_a)
(L^{\sigma_j}\otimes L^{\nu_j})^{\dagger}=
\alpha\frac{I\otimes I^{phys}+F\otimes F^{phys}}{2}
+\beta\frac{I\otimes I^{phys}-F\otimes F^{phys}}{2}
\eneq
\beq
(UI)^{\otimes 2}I(UI)^{\dagger \otimes 2}=
I
\eneq
\beq
(L^{\sigma_j}\otimes L^{\nu_j})I(L^{\sigma_j}\otimes L^{\nu_j})^{\dagger}=
I\otimes I^{phys},
\eneq
and
\beq
(UR)^{\otimes 2}F(UR)^{\dagger \otimes 2}=
\alpha \Pi_s - \beta \Pi_a
\eneq
\beq
(L^{\sigma_j}\otimes L^{\nu_j})(\alpha \Pi_s - \beta \Pi_a)(L^{\sigma_j}\otimes L^{\nu_j})^{\dagger}=
\alpha\frac{I\otimes I^{phys}+F\otimes F^{phys}}{2}
-\beta\frac{I\otimes I^{phys}-F\otimes F^{phys}}{2},
\eneq
\beq
(UI)^{\otimes 2}F(UI)^{\dagger \otimes 2}=
F
\eneq
\beq
(L^{\sigma_j}\otimes L^{\nu_j})F(L^{\sigma_j}\otimes L^{\nu_j})^{\dagger}=
F\otimes F^{phys}.
\eneq
Putting everything together
\beq
(L^{\sigma_j}\otimes L^{\nu_j})(UR)^{\otimes 2}I(UR)^{\dagger \otimes 2}(L^{\sigma_j}\otimes L^{\nu_j})^\dagger=
\alpha\frac{I\otimes I^{phys}+F\otimes F^{phys}}{2}
+\beta\frac{I\otimes I^{phys}-F\otimes F^{phys}}{2}
\eneq
\beq
(L^{\sigma_j}\otimes L^{\nu_j})(UR)^{\otimes 2}F(UR)^{\dagger \otimes 2}(L^{\sigma_j}\otimes L^{\nu_j})^\dagger=
\alpha\frac{I\otimes I^{phys}+F\otimes F^{phys}}{2}
-\beta\frac{I\otimes I^{phys}-F\otimes F^{phys}}{2},
\eneq
and
\beq
(L^{\sigma_j}\otimes L^{\nu_j})(UI)^{\otimes 2}I(UI)^{\dagger \otimes 2}(L^{\sigma_j}\otimes L^{\nu_j})^\dagger=
I\otimes I^{phys}\eneq
\beq
(L^{\sigma_j}\otimes L^{\nu_j})(UI)^{\otimes 2}F(UI)^{\dagger \otimes 2}(L^{\sigma_j}\otimes L^{\nu_j})^\dagger=
F\otimes F^{phys}
\eneq
from which one can also derive
\beqy
&&(L^{\sigma_j}\otimes L^{\nu_j})(UR)^{\otimes 2}(I\otimes I^{phys}+F\otimes F^{phys})(UR)^{\dagger \otimes 2}(L^{\sigma_j}\otimes L^{\nu_j})^\dagger=\\
&&\alpha (I \otimes I^{phys} + F \otimes F^{phys})\otimes \Pi_s^{phys}
+\beta (I\otimes I^{phys} - F \otimes F^{phys})\otimes \Pi_s^{phys}
\eneqy
\beqy
&&(L^{\sigma_j}\otimes L^{\nu_j})(UR)^{\otimes 2}(I\otimes I^{phys}-F\otimes F^{phys})(UR)^{\dagger \otimes 2}(L^{\sigma_j}\otimes L^{\nu_j})^\dagger=\\
&&\alpha (I \otimes I^{phys} + F \otimes F^{phys})\otimes \Pi_s^{phys}
+\beta (I \otimes I^{phys} - F \otimes F^{phys})\otimes \Pi_s^{phys}.
\eneqy
For the left boundary sites, where $R=I$, we have
\beqy
&&(L^{\sigma_j}\otimes L^{\nu_j})(UI)^{\otimes 2}(I\otimes I^{phys}+F\otimes F^{phys})(UI)^{\dagger \otimes 2}(L^{\sigma_j}\otimes L^{\nu_j})^\dagger=\\
&&I\otimes I^{phys} \otimes I^{phys} + F\otimes F^{phys} \otimes F^{phys},
\eneqy
\beqy
&&(L^{\sigma_j}\otimes L^{\nu_j})(UI)^{\otimes 2}(I\otimes I^{phys}-F\otimes F^{phys})(UI)^{\dagger \otimes 2}(L^{\sigma_j}\otimes L^{\nu_j})^\dagger=\\
&&I\otimes I^{phys} \otimes I^{phys} - F\otimes F^{phys} \otimes F^{phys}.
\eneqy
Using the above equations the updated $\rho_R$ state can then be written as
\beq
\rho_R=\alpha
\frac{(I \otimes I^{phys} +F \otimes F^{phys})\otimes \Pi_s^{phys}}{2d^2 D_s^{\chi}}
+\beta
\frac{(I \otimes I^{phys} -F \otimes F^{phys})\otimes \Pi_a^{phys}}{2d^2 D_a^{\chi}}.
\eneq
For a chain with only three qubits we then would have the following result for the second moment state
\beq
\rho^{(3)}=\alpha\frac{\Pi_{s,d^2}^{phys}\otimes \Pi_{s,d}^{phys}}{d^2 D_s^{\chi}}
+\beta\frac{\Pi_{a,d^2}^{phys}\otimes \Pi_{a,d}^{phys}}{d^2 D_a^{\chi}},
\eneq
where the subscript of the projector denotes both the symmetric or anti-symmetric character and the square root dimension of the Hilbert space they act on. The analogous calculation with an additional qubit would provide
\beqy
\rho^{(4)}&=&
\frac{\alpha\alpha}{d^2 D^{\chi}_s}\Pi_{s,d^2}^{phys}\otimes\Pi_{s,d}^{phys}\otimes \Pi_{s,d}^{phys}\\
&+&\frac{\beta\alpha}{d^2 D^{\chi}_s}\Pi_{a,d^2}^{phys}\otimes\Pi_{a,d}^{phys}\otimes \Pi_{s,d}^{phys}\\
&+&\frac{\alpha\beta}{d^2 D^{\chi}_a}\Pi_{s,d^2}^{phys}\otimes\Pi_{a,d}^{phys}\otimes \Pi_{a,d}^{phys}\\
&+&\frac{\beta\beta}{d^2 D^{\chi}_a}\Pi_{a,d^2}^{phys}\otimes\Pi_{s,d}^{phys}\otimes \Pi_{a,d}^{phys}.
\eneqy
We now have all the ingredients to evaluate the second moment state in the case of $N$ qubits, and we can write it compactly as follows (we drop the superscript in the projector operators since it is now irrelevant)
\beq
\aveU{\proj{\psi}^{\otimes 2}}=
\sum_{\boldsymbol{p}\backslash p_{L+1}}^{\{s,a\}^{N-L-1}} 
\frac{c_{p_L}c_{p_{L+2}}\cdots c_{p_N}}{d^2D_{p_N}^\chi}
\Pi_{p_L,d^L}\otimes\Pi_{p_{L+1},d}\otimes \cdots \otimes\Pi_{p_N,d},
\eneq
where $\boldsymbol{p}=\{p_L,p_{L+1},\dots,p_N\}$, and the sum runs over all $p_j=\{s,a\}$ expect for $p_{L+1}$. The subscript $p_{L+1}$ is $s$ or $a$ depending on the global parity provided by the other indexes $\{p_L,p_{L+2},\dots,p_N\}$ (if there is an odd number of antisymmetric projector then $p_{L+1}=a$, otherwise $p_{L+1}=s$); while the c coefficients are defined as $c_s=\alpha$ and $c_a=\beta$. Note that the above expression differs in general from the second moment states of Haar-distributed random pure state because it also involves anti-symmetric projectors. In the following section we provide a perturbative expression (in the bond-dimension) for the MPS second moment state which shows the relation with respect to the second moment Haar state. 

\subsection{Perturbative expression}
We start by providing a perturbative expression in $\chi^{-1}$ of the parameters $\alpha, \beta, D^\chi_{s,a}$, which can all be easily checked to satisfy the following equalities
\beqynn
\frac{1}{D^\chi_s}&=&\frac{2}{\chi(\chi+1)}
=\frac{2}{\chi^2}\left(1+\frac{1}{\chi}\right)^{-1}
=\frac{2}{\chi^2}\left[1-\frac{1}{\chi}+O\left(\chi^{-2}\right)\right]=\frac{2}{\chi^2}+O\left(\chi^{-3}\right)\\
\frac{1}{D^\chi_a}&=&\frac{2}{\chi(\chi-1)}
=\frac{2}{\chi^2}\left(1-\frac{1}{\chi}\right)^{-1}
=\frac{2}{\chi^2}\left[1+\frac{1}{\chi}+O\left(\chi^{-2}\right)\right]=\frac{2}{\chi^2}+O\left(\chi^{-3}\right)\\
\alpha&=&\frac{\chi+1}{2(2\chi+1)}
=\frac{1}{4}\left(1+\frac{1}{2\chi}\right)^{-1}+\frac{1}{4\chi}\left(1+\frac{1}{2\chi}\right)^{-1}=\frac{1}{4}+\frac{1}{8\chi}+O\left(\chi^{-2}\right)\\
\beta&=&\frac{\chi-1}{2(2\chi-1)}
=\frac{1}{4}\left(1-\frac{1}{2\chi}\right)^{-1}-\frac{1}{4\chi}\left(1-\frac{1}{2\chi}\right)^{-1}=\frac{1}{4}-\frac{1}{8\chi}+O\left(\chi^{-2}\right).\\
\eneqynn
Using the above expansions and the previous calculation, the perturbative expression of the second moment state after averaging over the last two sites unitaries provides
\beq
\frac{(I+F)\otimes \Pi_s^{phys}}{2d^2 D_s^{\chi}}
+\frac{(I-F)\otimes \Pi_a^{phys}}{2d^2 D_a^{\chi}}
=
\frac{2}{d^2\chi^2}\frac{I\otimes I_d^{phys}+F\otimes F_d^{phys}}{2}+O(\chi^{-3}),
\eneq
meaning that the operator norm of the sub-leading terms are at least $\chi$ times smaller than the operator norm of the leading term. Evaluating the average with respect to the next qubit we have
\beq
\frac{1}{2d^2\chi^2}
\left[
\frac{I\otimes I_d^{phys}\otimes I_d^{phys}+F\otimes F_d^{phys}\otimes F_d^{phys}}{2}
+O(\chi^{-1})
\right].
\eneq
Note that the leading term is just the projector over the symmetric subspace, and the iterative calculation would provide for a chain of $N$ qubits the following expression
\beq\label{eq:pertu}
\aveU{\psi^{\otimes 2}}=\frac{2}{d^2\chi^2}\left(\frac{1}{4}\right)^{N-1-\log_d{\chi}}
\left[\Pi_{s,D}^{phys}+O\left(\chi^{-1}\right)\right]
=\frac{2}{D^2}\left[\Pi_{s,D}^{phys}+O\left(\chi^{-1}\right)\right].
\eneq 
Hence to leading order in $\chi^{-1}$ the MPS second moment state is exponentially close, in the size of the system, to the second moment Haar state $\frac{2}{D(D+1)}\Pi_{s,D}^{phys}$.  

\subsection{Statistical fluctuations}
In this section we use the previously obtained perturbative expression for the second moment to bound the statistical fluctuations in the estimation of finite temperature expectation values
\beq
\qexps{B}_T\equiv\frac{\Tr{\rho_T B}}{\Tr{\rho_T}}
=\frac{\aveU{\qexpl{\psi}{\sqrt{\rho_T}B\sqrt{\rho_T}}{\psi}}}{\aveU{\qexpl{\psi}{\rho_T}{\psi}}}.
\eneq
Defining $A\equiv\sqrt{\rho_T}$, the estimate of  $\qexps{B}_T$ is given sampling from random MPS the quantity 
\beq
z=\frac{\qexpl{\psi}{ABA}{\psi}}{\qexpl{\psi}{A^2}{\psi}}\equiv \frac{x}{y},
\eneq
which is the ratio of two correlated random variables $x$ and $y$. We denote with $\ave{x}$ the average of $x$ with respect to $\ket{\psi}$, and we define $\delta x=x-\ave{x}$ (and similarly for $y$ and $z$). Using standard error propagation we can write
\beq
\ave{z}\approx \frac{\ave{x}}{\ave{y}}-\frac{\ave{\delta x \delta y}}{\ave{y}^2}+\frac{\ave{x}\ave{\delta y^2}}{\ave{y}^3}.
\eneq
Since $\frac{\ave{x}}{\ave{y}}=\qexps{B}_T$ is the actual value that we want to estimate (using the result on the average random MPS state in Eq.\ref{eq:avestate}), but $\ave{z}$ is the outcome of our MC algorithm, we need to upper-bound the distance $\Big{\vert} \ave{z}-\frac{\ave{x}}{\ave{y}} \Big{\vert}$ between the two. We then have to estimate $\Big{\vert} \frac{\ave{\delta x \delta y}}{\ave{y}^2} \Big{\vert}$ and $\Big{\vert} \frac{\ave{x}\ave{\delta y^2}}{\ave{y}^3} \Big{\vert}$, which can be done as follows using the previously derived perturbative expression for the second moment state in Eq.\ref{eq:pertu}, and the well known identity $\Tr[F(X\otimes Y)]=\Tr[XY]$
\beq
\Big{\vert} \frac{\ave{\delta x \delta y}}{\ave{y}^2} \Big{\vert}
=\Big{\vert}
\frac
{\Tr\left[{\left(D^2\aveU{\psi^{\otimes 2}} -{I}\right)ABA\otimes A^2}\right]}
{{\Tr (A^2)}^2}
\Big{\vert}
\sim 
O(D^{-1})+O\left(\chi^{-1}\right)\frac{\ave{x}}{\ave{y}}
\eneq

\beq
\Big{\vert} \frac{\ave{x}\ave{\delta y^2}}{\ave{y}} \Big{\vert}
=
\Big{\vert}
\frac
{{\Tr (A^2B)}
\Tr\left[{\left(D^2\aveU{\psi^{\otimes 2}} -{I}\right)A^2\otimes A^2}\right]}
{{\Tr (A^2)}^3}
\Big{\vert}
\sim 
O(D^{-1})+O\left(\chi^{-1}\right)\frac{\ave{x}}{\ave{y}},
\eneq 
implying that the relative error in approximating the thermal average $\qexps{B}_T=\frac{\aveU{\qexpl{\psi}{ABA}{\psi}}}{\aveU{\qexpl{\psi}{A^2}{\psi}}}$ with $\aveU{\frac{\qexpl{\psi}{ABA}{\psi}}{\qexpl{\psi}{A^2}{\psi}}}$ is at most of the order of $\chi^{-1}$ plus terms exponentially small in $D$, the dimension of the Hilbert space. 

We can also derive an upper-bound on the relative variance of $z$ again simply using standard error propagation and linear algebra, together with Eq.\ref{eq:pertu}
\beqy
\Big{\vert}\frac{\ave{\delta z^2}}{\ave{z}^2}\Big{\vert}
&\approx&
\Big{\vert}
\frac{\ave{\delta x^2}}{\ave{x}^2}
+\frac{\ave{\delta y^2}}{\ave{y}^2}
-2\frac{\ave{\delta x \delta y}}{\ave{x}\ave{y}}
\Big{\vert}\\
&\leq &
\Big{\vert}\frac{\Tr C^2}{\Tr{C^{\otimes 2}}}\Big{\vert}
+\Big{\vert}\frac{\Tr A^4}{\Tr(A^2)^{\otimes 2}}\Big{\vert}
+2\Big{\vert}\frac{\Tr(A^2C)}{\Tr{A^2\otimes C}}\Big{\vert}+O\left(\chi^{-1}\right)\\
&\sim &
O(D^{-1})+O(\chi^{-1}),
\eneqy 
where $C\equiv A^2B$. This last result can then be plugged in the Chebyshev's inequality to derive the $(\epsilon,\delta)$-approximation scheme in the main manuscript. 

\end{document}